\documentclass[]{spie}  
\usepackage[]{graphicx}

\title{Disks around young stars with VLTI/MIDI}

\author{
Roy~van~Boekel\supit{a}, 
P\'eter~\'Abrah\'am\supit{b}, 
Serge~Correia\supit{c},
Alex~de~Koter\supit{d}, 
Carsten~Dominik\supit{d}, 
Anne~Dutrey\supit{e}, 
Thomas~Henning\supit{a}, 
\'Agnes~K\'osp\'al\supit{b},
R\'egis~Lachaume\supit{f},
Christoph~Leinert\supit{a}, 
Hendrik~Linz\supit{a}, 
Michiel~Min\supit{d}, 
L\'aszl\'o~Mosoni\supit{b},
Thomas~Preibisch\supit{g}, 
Sascha~Quanz\supit{a}, 
Thorsten~Ratzka\supit{a}, 
Alexander~Schegerer\supit{a}, 
Rens~Waters\supit{d}, 
Sebastian~Wolf\supit{a},
and Hans~Zinnecker\supit{c}
\skiplinehalf
\supit{a}Max-Planck Institut f\"ur Astronomie, K\"onigstuhl 17, D-69117 Heidelberg, Germany; \\
\supit{b}Konkoly Observatory, Hungarian Academy of Sciences, PO Box 67, 1526 budapest, Hungary; \\
\supit{c}Astrophysikalisches Institut, An der Sternwarte 16, D-14482 Potsdam, Germany. \\
\supit{d}University of Amsterdam, Kruislaan 403, 1098 SJ Amsterdam, Netherlands; \\
\supit{e}Observatoire de Bordeaux, 2 rue de l'Observatoire, F-33270 Floirac, France; \\
\supit{f}CRyA UNAM, Antigua Carretera a P\'atzcuaro 8701, Morelia, Michoac\'an, M\'exico; \\
\supit{g}Max-Planck-Institut f\"ur Radioastronomie, Auf dem H\"ugel 69, D-53121 Bonn, Germany.
}

\begin{document} 
 \maketitle 

\begin{abstract}

We report on observations of circumstellar disks around young stars that have been obtained with the MIDI instrument, which is mounted on the VLT Interferometer and operates in the 10\,$\mu$m atmospheric window. The maximum spatial resolution of 5~milli-arcsec corresponds to sub-AU scales at the distance to nearby star formation regions. Thus, we can study the disks on the spatial scales at which important processes occur, such as accretion, dust processing, and planet formation. The main results obtained so far can be summarized as follows: 1. The measured interferometric visibilities are in good qualitative agreement with those predicted by models of circumstellar disks. In particular, a predicted correlation between the strength of the far-infrared excess and the spatial structure of the disk is confirmed by direct measurements; 2. In several objects strong evidence for deviations from circular symmetry is present, indicating that an inclined disk is indeed the dominant component seen in the mid-infrared; 3. The dust properties are not uniform over the disk, but are instead a strong function of distance to the central star. The dust in the innermost disk regions is observed to be more ``processed'' than the dust further out, both in Herbig~Ae star disks and in those around T\,Tauri stars.

%



\end{abstract}



\section{Introduction}
\label{sec:intro}
We report on observations of young stars that have been obtained with MIDI since the first successful measurements were performed in June 2003. First we briefly introduce the MIDI instrument and its main scientific objectives, followed by an introduction to star formation. In sections~\ref{sec:low_mass}, \ref{sec:intermediate_mass} and \ref{sec:high_mass} we report results on young stars of low, intermediate, and high mass, respectively. In the last section, we take a look at future prospects.

\subsection{VLTI and MIDI}
\label{sec:instrument}
We start with a very brief description of the Very Large Telescope Interferometer (VLTI) and the MID Infrared instrument (MIDI). The complex VLTI infrastructure allows the light from VLT's four 8.2\,m Unit Telescopes (UTs) to be \emph{coherently} combined in the interferometric laboratory. Additionally, four movable 1.8\,m Auxiliary Telescopes (ATs) can be placed on 30 ``docking stations'', allowing virtually any baseline configuration (length and orientation) to be realized. The longest baselines ($\sim$200\,m) provide a spatial resolution of about 5\,mas ($\lambda/2B$, where $\lambda$ is the observing wavelength and $B$ the baseline length) at 10\,$\mu$m. MIDI is a classical Michelson interferometer capable of combining the light from two telescopes at a time. It allows the measurement of interferometric fringes in a number of filters, but is usually used in spectrally dispersed mode. As dispersive elements a prism (spectral resolution $R$$\sim$$30$) and a grism ($R$$\sim$$230$) are available. The visibility accuracy depends on the observing mode and the atmospheric quality of the night. The goal is to reach 1\,\% RMS in the most accurate mode during a good night. The observations presented here were taken in a less accurate (but more sensitive) observing mode at a wide range of observing conditions, and typically have visibility accuracies of 5 to 15\%. The detection limits are roughly 200\,mJy \emph{correlated} flux\footnote{$F_{c}=F_{t}*V$, where $F_{c}$ and $F_{t}$ are the correlated and total source flux, respectively, and $V$ is the (absolute value of) the visibility.} for the UTs without external fringe tracking during good conditions, and roughly 10\,Jy for the ATs. When the external fringe tracker FINITO becomes available, it is anticipated that these limits will improve by 4 to 5 magnitudes. Note that all Unit Telescopes are equipped with an Adaptive Optics system, ensuring diffraction limited beams entering MIDI at virtually all times. For a more detailed description of MIDI and VLTI we refer to Leinert et al.\cite{2003SPIE.4838..893L} and Sch\"oller et al. (this volume).

\subsubsection{MIDI science}
\label{sect:midi_science}
MIDI has been built to operate in the 10$\mu$m atmospheric window (N-band), and could therefore in principle be used to observe any object that radiates in this wavelength region. In practice however, sources need to fulfill two criteria to be suitable sources for MIDI: 1. they need to be sufficiently bright, and 2. the spatial scale of the emission should be in the "interesting" range for MIDI (approximately 0.2 to 0.005 arcseconds: larger objects are spatially resolved by the current generation of 8-10m telescopes, smaller objects remain unresolved even on the longest available baselines).

Naturally, regions where warm/hot dust ($\sim$200 to 1500 K) is present best fulfill these criteria, and the scientific requirements for the study of such environments have been the defining factor in the instrument design. Specifically, the capability of measuring \emph{spectrally resolved visibilities} enables detailed studies into the nature of the dust. The two most prominent fields of research that benefit from MIDI's capabilities are the study of circumstellar environments, and of dust tori in Active Galactic Nuclei (AGN). Here, the former can be divided into the study of dust around Young Stellar Objects (YSOs), and of dust around evolved stars. Dust around main-sequence stars usually contributes significantly less than 1\% to the total flux of such systems in the 10\,$\mu$m spectral region. 
Thus, extra-zodiacal dust clouds cannot be studied with MIDI. 

Here, we will focus on observations of the dust around newly formed stars. For an overview of MIDI results on evolved stars and AGN we refer to Ratzka et al. (this volume) and references therein.
In all the above mentioned fields, the study of dusty environments essentially boils down to answering two broad questions:

1. What is the spatial distribution of the warm/hot dust?

2. What is the composition of this dust?

\noindent

\subsection{Star formation}
\label{sec:star_formation_intro}

Stars form out of the collapse of large clouds of cold, relatively high density gas and dust that exist in the interstellar medium and are known as \emph{molecular clouds}. Many stars and their associated planetary systems can form from a single cloud by fragmentation. In the study of star formation a crude division is usually made between between a "low mass" and a "high mass" regime. Though many --~possibly most~-- stars form in cluster environments, we will describe here how \emph{isolated} low mass stars are believed to form. Most nearby young stars that were studied with MIDI so far and are reported on here, formed in isolation. Star formation in dense clusters is believed to be different in some respects which we do not discuss here\cite{2004RvMP...76..125M}.


   \begin{figure}
\vspace{-0.1cm}
   \begin{center}
   \begin{tabular}{c}
   \includegraphics[height=11.6cm,width=10cm]{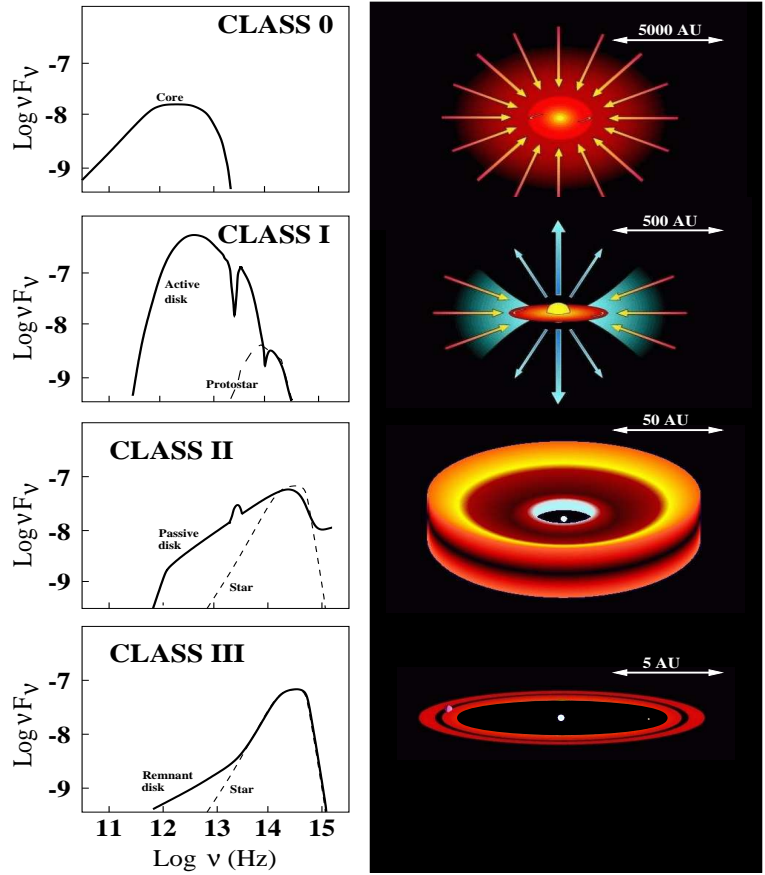}
   \end{tabular}
   \end{center}
\vspace{-0.2cm}
   \caption[example]
   { \label{fig:classes}
A schematic overview of the four phases distinguished in low mass star formation (see section~\ref{sec:star_formation_intro}). On the right, typical spectral energy distributions of sources in each phase a shown. On the right, the geometry of the system is sketched. Figure courtesy of Mark McCaughrean, Antonella Natta and Vincent Icke, adapted with permission.
}
   \end{figure}



The process of \emph{low mass star formation} can schematically be divided into four phases, covering the evolution from a collapsing cloud to full grown star, possibly with a developed planetary system. These phases can observationally be distinguished by their spectral energy distribution (SED), and sources in the respective phases are identified as class~0 through III in the Lada classification\cite{1984ApJ...287..610L,1987ApJ...312..788A,1993ApJ...406..122A}. In figure~\ref{fig:classes} we show sketches of sources in each of the four phases, and the corresponding SEDs.
In the first phase (observationally class~0), the density at the core of a collapsing molecular cloud fragment increases rapidly as material falls in more or less spherically. Due to rotation and conservation of angular momentum, a flattened structure builds around the forming star, which will become the circumstellar disk as it flattens further. The central object itself is in this phase called a \emph{proto}-star since nuclear fusion does not yet take place in its center. Class~0 sources can be observed at far-infrared and millimeter wavelengths only. 
In the second phase (class~I), the central object has become much more compact and the angular momentum of the cloud material inhibits further spherical infall. Accretion onto the forming star proceeds through the disk while the outer disk regions continue to be supplied with "fresh" material from the surrounding cloud. In the direction perpendicular to the disk, part of the accreting material may be ejected from the system in a bi-polar outflow, the rest is accreted by the (proto-)star. Class~I sources are very bright in the infrared, and the main energy source for this radiation is the release of gravitational energy by accreting material in the disk. Hence, such accreting disks are called \emph{active} disks.
The third phase (class~II) begins when the supply of fresh material in the outer regions of the disk comes to a halt. The forming star has essentially reached its final mass, but is still contracting and will yet become significantly smaller and hotter. The regions above and below the disk do no longer contain much obscuring dust, and the star is optically visible. The disk is still very bright and dominates the SED at infrared and  millimeter wavelengths. However, in this phase the main energy source in the disk is absorption of stellar radiation and the disk is said to be \emph{passive}. The disk is initially still massive but will disperse on a time scale of several million years. It is during this phase that planets are believed to form, and most of the observations presented here are of sources in this evolutionary stage. Low mass stars in the class~I and class~II phase are called T\,Tauri stars.
In the fourth and final phase there is only a gas-poor \emph{remnant} disk left. Giant gas planets no longer form but terrestrial planets may continue to assemble through the merging of larger bodies ("planetesimals") made of refractory material. The $\beta$~Pic system is a famous example of such a \emph{class~III} source.

\newpage
\emph{High mass stars} form on much shorter time scales and are deeply enshrouded in molecular cloud material during their entire pre-main sequence evolution. When their intense UV radiation fields have cleared the surroundings and the stars become optically visible, they are already on the main sequence. It is unknown to what degree their formation is a scaled-up and speeded-up version of the low mass scenario described above. High mass stars may form in a qualitatively different way from their low mass cousins. Recent evidence shows that also high mass stars are surrounded by flattened circumstellar structures with scales of several thousand~AU, possibly accretion disks, in early stages of their evolution \cite{
2005Natur.437..112J,2006ApJ...637L.129S}. Nonetheless it is still unclear whether the circumstellar emission is disk or envelope dominated, and how similar massive star disks are to those around low mass YSOs. Specifically, high mass star ``disks'' are very large and have been studied only on these large scales. It is unknown what the disks look like on much smaller scales. As high mass stars are rare, and their disks are short-lived due to photo-evaporation\cite{2000prpl.conf..401H}, we typically find them only at large distances.

Herbig Ae/Be stars \cite{1960ApJS....4..337H} are young stars in the range of 2 to 8 M$_{\odot}$ and are traditionally considered a separate class of \emph{intermediate mass} YSOs. Stars up to 3 or 4\,M$_{\odot}$ are called Herbig~Ae (HAE) stars, and there is growing evidence that their disks bear great similarity in terms of their properties and evolution with those of lower mass stars, and that this similarity extends well into the Brown Dwarf regime \cite{2005Sci...310..834A}. Thus, the division between T\,Tauri stars and HAE stars may be artificial. To be consistent with existing literature we choose to present MIDI observations of Herbig stars (all of which concern HAE stars) in a separate ``intermediate mass'' section. It is yet unclear whether the more massive members of the class, the Herbig Be (HBE) stars, bear more overall resemblance to low or to high mass YSOs. This requires further study.


\vspace{0.1cm}

\noindent
\emph{Dust in circumstellar disks} \\
The main ingredient for stars and planets - and the disks in which they form - is gas. The dust, though unimportant as a mass contributor (gas/dust ratios of 100 by mass are usually assumed), plays a vital role in the whole formation process of stars and their associated planetary systems. Unlike the gas, the dust can efficiently radiate at low temperatures. In the early collapse phase this allows the material to cool, and without the dust the Jeans mass is predicted to be so high that only very massive stars would form\cite{2002Sci...295...93A}. The cooling and heating of disk material, and thus its temperature, is governed by the dust properties. The temperature in turn determines the pressure scale height and thus the spatial structure of the disk. A fortunate side effect of the large opacity of the dust, though not of immediate concern to the disks themselves, is that it allows us to study these objects at a wide range of wavelengths and spatial scales. More importantly, dust grains involved in low velocity collisions can stick together and form larger aggregates, which may meet with other aggregates to form yet larger structures, and so forth. At some point, the largest structures will be massive enough to have a significant gravitational potential, and these bodies may develop into planets.

\begin{figure}
   \begin{center}
   \begin{tabular}{c}
   \includegraphics[height=5cm]{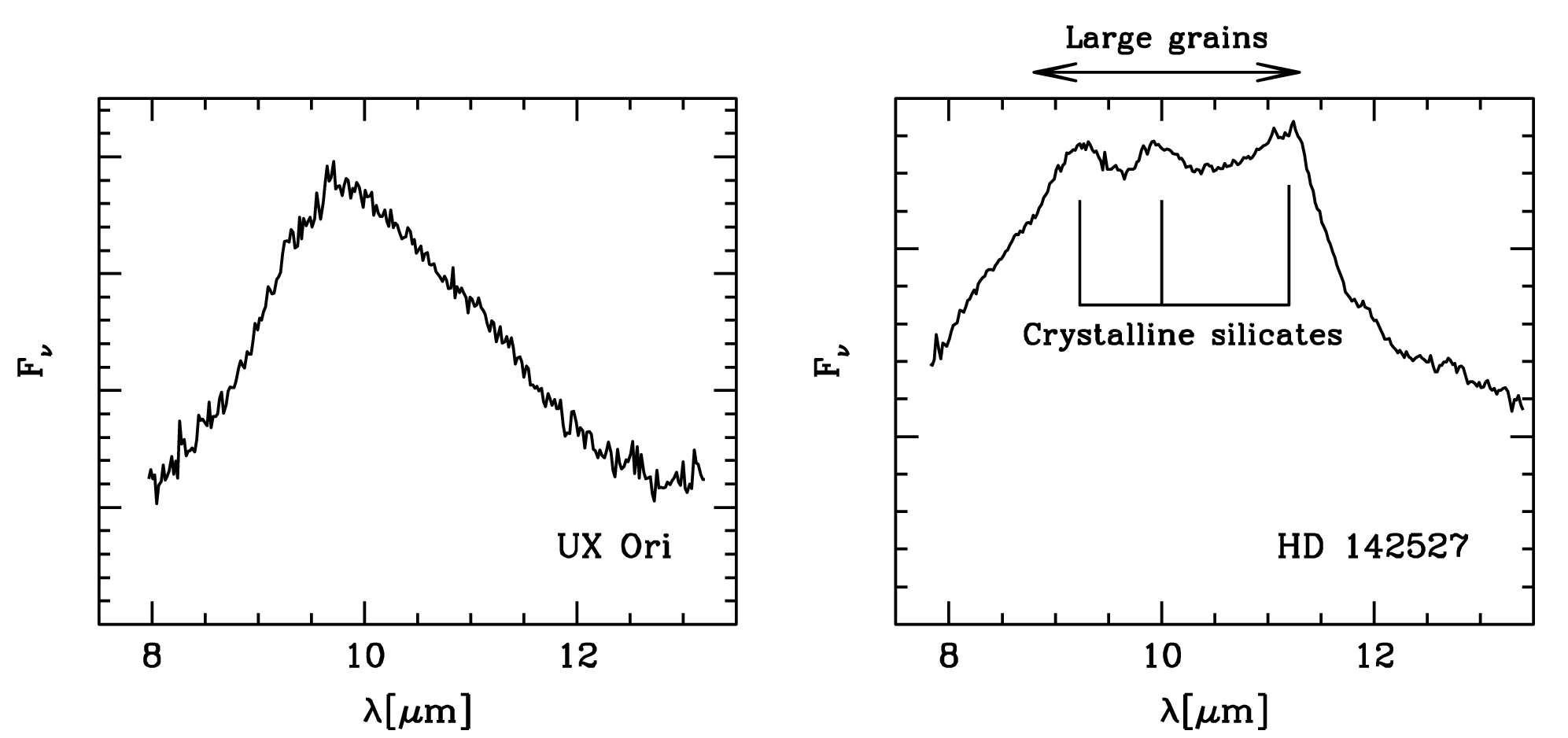}
   \end{tabular}
   \end{center}
   \caption[example]
   { \label{fig:processing}
The effects of dust processing on the 10\,$\mu$m silicate feature. The left spectrum is typical of small, amorphous grains as they are found in the ISM. In the right spectrum the broader, flat-topped silicate band implies grain growth, whereas the additional narrower bands witness the presence of crystalline material.
}
   \end{figure}

The dust in molecular clouds and circumstellar disks consists mainly of carbon and silicates. Whereas the opacity of carbon does not show much spectral structure in the infrared, small silicate grains have strong resonances. A very prominent spectral feature is present in the 10~micron atmospheric window in which MIDI operates. This \emph{"10~micron feature"} is seen in emission towards many young stars but may also be seen in absorption in very embedded sources or disks seen edge-on (in general:  in situations where we see a warm background through a large column of cold dust). The appearance of the 10~micron feature depends on chemical composition, lattice structure, and grain size of the silicates and can therefore be used to determine the dust properties. The spectra in figure~\ref{fig:processing} illustrate the diagnostic power of the 10~micron silicate feature for the evolutionary state of the dust. The emission band on the left is reminiscent of sub-micron sized amorphous grains, typical for the interstellar medium (ISM) and the supposed original ingredient of a circumstellar disk. The 10~micron feature of such ``pristine'' dust is triangularly shaped, peaking around 9.7\,$\mu$m. The spectrum on the right shows a much broader, flat-topped emission band. This is mainly due to \emph{growth} of the dust grains to several microns. A second process altering the dust is \emph{crystallization} of the initially amorphous material, which occurs whenever the dust reaches temperatures above $\sim$900\,K. Contrary to grain growth, which can occur essentially anywhere in the disk due to the relatively high densities, crystallization is a process that requires much more ``special'' circumstances (high temperatures). Crystalline silicates give rise to additional, narrow emission bands. These are observed in many young stars and also in solar system comets. The origin of the crystalline silicates in comets is a matter of debate. Surely, the high temperatures needed for their production prevailed in the innermost disk regions, both during the passive disk phase in which we observe the crystals and in the active disk phase that preceded it. However, the comets have formed much further out, and their formation zone has been frozen during the entire evolution of the solar nebula. Yet, we find crystalline silicates in comets, sometimes in large abundances.  The two competing theories to explain this can be summarized as follows: 1. the crystals formed in the innermost disk regions, and were transported outward by radial mixing processes before being incorporated in the comets;  and 2. the crystals formed in the comet formation zone itself in transient heating events, where material was briefly heated to the required temperatures in shocks\cite{2002ApJ...565L.109H} or lightning bolts\cite{1998A&A...331..121P,2000Icar..143...87D }. Dust displaying a 10~micron feature with strong signs of processing is also called ``evolved''.

\vspace{0.1cm}
\noindent
\emph{The potential of MIDI for circumstellar disk studies} \\
Important processes in circumstellar disks include accretion, dust processing, possibly radial mixing and eventually planet formation. The relevant spatial scales range from an AU or less for the dominant release of accretion energy and thermal dust processing in the innermost disk regions, to typically some 10\,AU for giant planet formation. Given the typical distance to nearby star formation regions of 100\,pc or more, such scales are too small to be spatially resolved in the infrared for the current generation of large (8-10\,m class) telescopes. The maximum spatial resolution of MIDI of 5\,milli-arcsec corresponds to 0.5\,AU at a distance of 100\,pc. Thus, with VLTI/MIDI we are well equipped to study the relevant processes in circumstellar disks on the scales at which they occur. This allows us to address very specific questions related to the spatial distribution and properties of the dust in circumstellar disks. Is the observed variety in spectral energy distributions due to differences in disk geometry (see also section~\ref{sec:leinert})? Are the "puffed-up" inner rim and the relatively cool ``shadowed'' region behind it, predicted by recent theoretical models \cite{2001A&A...371..186N,2001ApJ...560..957D}, really present? How do the dust properties change throughout the disk? Specifically: is the efficiency of grain growth a function of distance to the central star, and does the abundance of crystalline silicates change with distance to the central star in a way that is consistent with radial mixing of material from the innermost disk regions outward?

%
%

\subsection{Visibilities and their interpretation}
One should be aware that the information contained in visibility measurements made with optical/infrared interferometers such as MIDI is not straightforward to interpret, since some \emph{a priori} knowledge of what the source looks like is required. As a two-element interferometer, MIDI measures visibility amplitudes. Due to the piston term of the atmospheric turbulence which is not corrected for by the AO systems, the visibility phase cannot be measured\footnote{Note that some information on the differential phase (between different wavelengths) can be extracted, see \cite{2004SPIE.5491..588T} for an example of this exercise.}. This means that the reconstruction of actual images from the measured visibilities through aperture synthesis imaging is in general not possible. Rather, the visibilities should be compared directly to those predicted by a physical model of the source under investigation. In most cases, one has visibilities measured on only a very limited number number of baselines\footnote{This is certainly true for MIDI measurements of YSOs. Due to the current sensitivity limits, most of these can only be observed well with the 8.2\,m UTs, and measuring a source's visibility at \emph{one} position in the uv plane takes 1~hour, including a measurement of a calibration star. This amounts to 2~hours of 8\,m time and thus getting good uv coverage is a very costly process for the time being. The situation will drastically change for the better when the fringe tracker (FINITO) becomes operational, such that the movable 1.8m ATs --~fully dedicated to interferometry~-- can be used to observe a wide range of YSOs. This will enable VLTI to deploy its full potential and we may expect a leap forward in the level of detail in which we can study a range of celestial sources, including the circumstellar material around young stars.}. In order to interpret such sparse data, it is important to have a physical model that has few free parameters and is constrained as much as possible by complementary observations such as the spectral energy distribution, polarimetric measurements, and possibly spatially resolved measurements at other wavelengths. The classic example of a problem where we fit a parameterized model to visibility data is the determination of a stellar diameter. Here we have a very good \emph{a priori} idea of what the source looks like (a simple uniform disk or a limb-darkened model) and a single parameter (the angular diameter\footnote{When measurements are available at high spatial frequencies, beyond the first "null" of the stars' visibility curve, the limb darkening profile can be directly measured. In this case the model has more than one parameter.}) that is to be fine tuned to match the interferometric measurement\cite{1921ApJ....53..249M}.

   \begin{figure}
   \begin{center}
   \begin{tabular}{cc}
   \includegraphics[height=5.7cm]{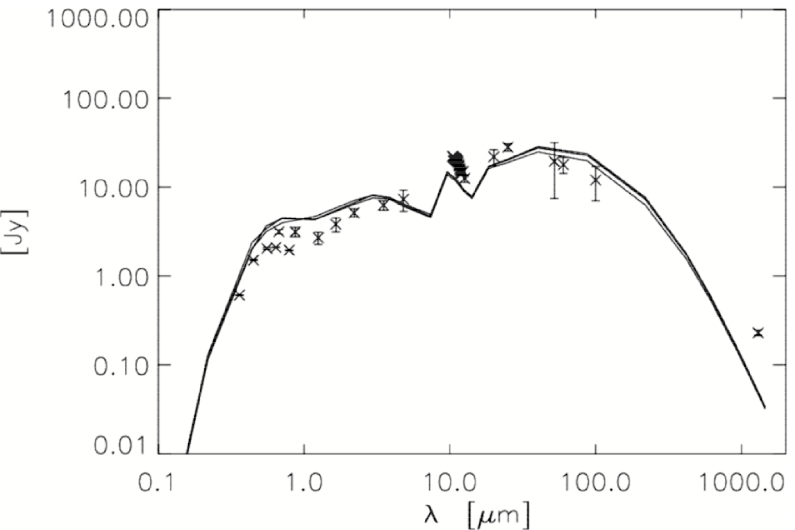} &
   \includegraphics[height=5.7cm]{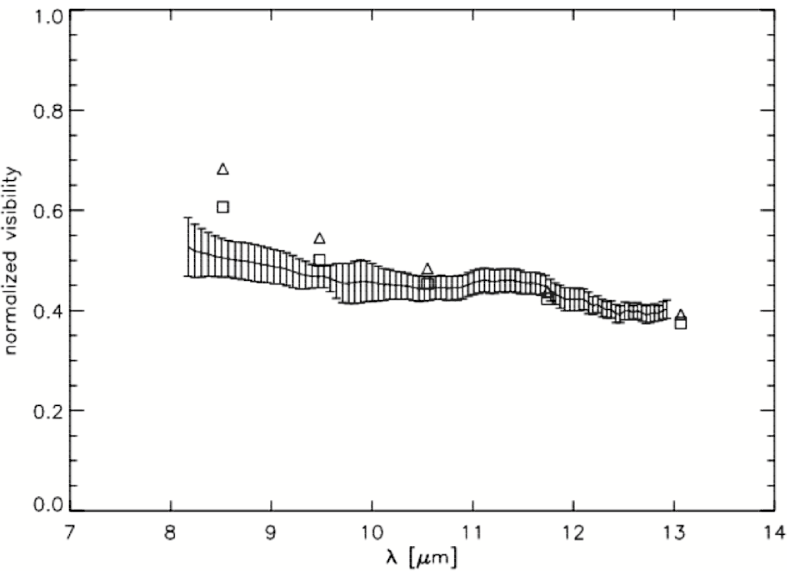}
   \end{tabular}
   \end{center}
\hspace{-0.4cm}
   \caption[example]
   { \label{fig:example}
A Radiative Transfer model fit to RY~Tau (see section~\ref{sec:schegerer}). \emph{Left} the observed and modeled spectral energy distribution. \emph{Right}: observed visibility on a baseline of 48\,m (curve with error bars), and the modeled visibility at 5 different wavelengths and in 2 perpendicular directions (squares for the disk major axis, triangles for the minor axis).
}
   \end{figure}

The dusty circumstellar environment of young stars is not anywhere nearly as well understood  theoretically and characterized observationally as the stellar photosphere in the above example. For low and intermediate mass YSOs the presence of circumstellar disks is firmly established (there is still ongoing discussion whether or not a spherical halo is present as an component \emph{additional} to the disk\cite{2003MNRAS.346.1151V}). For young high mass stars it is yet unclear whether the circumstellar environment is disk or envelope dominated. 

To model circumstellar disks, some authors use parameterized descriptions of the temperature, density and disk scale height as a function of distance to the central star. Typically these quantities are assumed to behave like power laws and the exponents are adjusted to simultaneously fit the observed visibilities and SED. Additionally, theoretical constraints (such as the disk surface density behaving like $\Sigma \propto R^{-3/2}$, 
\cite{1977Ap&SS..51..153W} or $\Sigma \propto R^{-1}$,\cite{1997ApJ...474..397D}) are often imposed to limit the number of free model parameters. Such simple models allow for a quick scanning of parameter space and exploration of degeneracies. More realistic disk models employ radiative transfer calculations that solve for the temperature structure of the disk. The density structure is either parameterized or calculated self-consistently together with the disk temperature in an iterative process where hydrostatic equilibrium is assumed in the vertical direction. These models incorporate much physics and have relatively few free parameters, thus offering powerful tools to interpret sparse visibility data. However, they are very demanding in terms of computational power and fine-tuning models to fit individual sources is a time consuming process. Exploring degeneracies in such modeling results presents an appreciable challenge.

\section{MIDI results: low mass stars} 
\label{sec:low_mass}

\subsection{The structure of T\,Tauri star disks}
In this section we present several examples of studies in which models fits are made to the spectral energy distribution and interferometric visibilities simultaneously. This approach is often useful for lifting degeneracies present in fits to either one separately.

\subsubsection{The disk of Ry~Tauri}
\label{sec:schegerer}

RY\,Tau is a bright T\,Tauri star with an age of approximately 6\,Myr\cite{1999A&A...342..480S} at a distance of 135\,pc\cite{1999A&A...352..574B}. It was was recently observed with the Mid-Infrared Interferometer (MIDI/VLTI) at projected baseline lengths of 48 and 78\,m. We have modeled the observations using a disk model in which the temperature and vertical disk structure are computed self-consistently, using the radiative transfer code MC3D\cite{1999A&A...349..839W}. The disk material is heated both by stellar radiation and an additional accretion component.  We could reproduce spectroscopically resolved visibility data from MIDI and the spectral energy distribution simultaneously. 

We found $T_\mathrm{\star}$=5560\,K, $L_\mathrm{\star}$=11.0\,L$_{\odot}$ and $R_\mathrm{\star}$=3.6\,R$_{\odot}$ as properties of the central star. The stellar temperature corresponds to a star of spectral type F8\,III. The modeling results for disk properties are: $R_\mathrm{in}$=0.2\,AU, $R_\mathrm{out}$=70\,AU, as well as $M_\mathrm{disk}$=0.025\,M$_{\odot}$. An accretion rate of 2.5e-7\,M$_{\odot}$/yr was determined. We assume an interstellar visual extinction of 2.7\cite{1990AJ.....99..924B}. Both spectroscopically resolved visibility points do not provide strong constraints on inclination or even the position angle (PA) of the disk. However, models with inclinations larger than 45$^{\circ}$ do not reproduce the SED and visibility at all. Our modeling results confirm the properties of RY\,Tau which were derived recently from near-infrared measurements using the Palomar Testbed Interferometer\cite{2005ApJ...622..440A}. It is an important result that interferometric observations in two different wavelength regimes reflect similar modeling results. Only the disk masses differ by a factor of $<$2. A publication reporting these results is in preparation (Schegerer et al., in prep.).

\subsubsection{FU~Orionis}
\label{sec:quanz}

   \begin{figure}
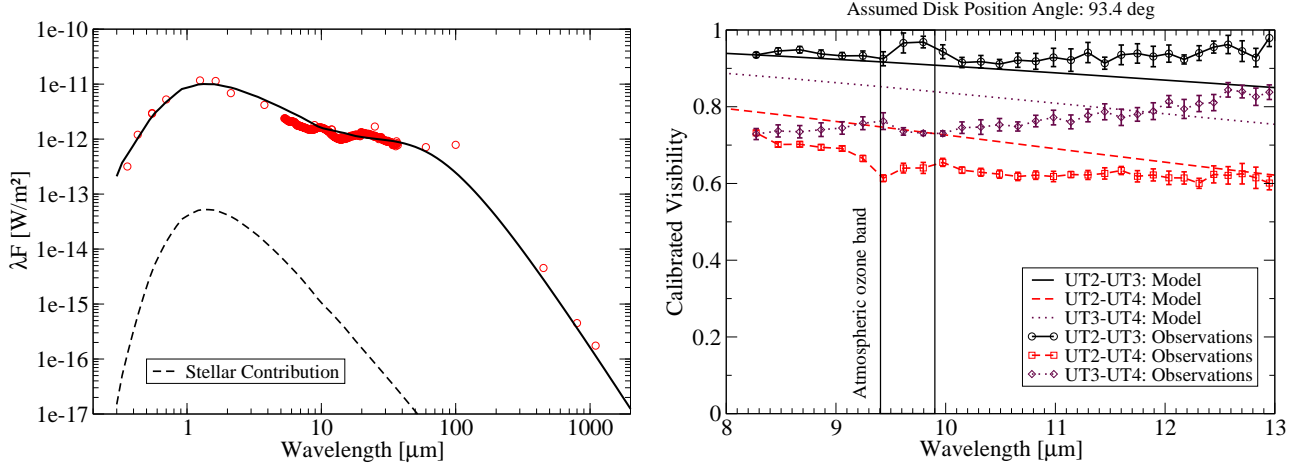

   \begin{center}
   \begin{tabular}{cc}
   \hspace{-0.25cm}
   \includegraphics[height=5.9cm]{FU_Ori_SED_Paper_new_55+109deg.eps} &
   \includegraphics[height=6.15cm]{f11b.eps}
   \end{tabular}
   \end{center}
   \caption[example]
   { \label{fig:quanz}
\emph{Left:} the SED of FU~Ori. The thin solid curve shows our disk model fit.
\emph{Right:} Observed and modeled visibilities of FU~Ori for three different baselines.
}
   \end{figure}

FU~Orionis objects (FUORs) are actively accreting (class~I in the Lada scheme) YSOs that are characterized by periods of highly increased accretion rates (outbursts), during which the luminosity can suddenly increase by a factor of $\sim 10^2$ and then diminishes again on time scales of ten to several hundred years. We observed the prototype of the class -~FU~Orionis itself~- with MIDI during three nights in October and November 2004. It was clearly resolved on two VLTI baselines and marginally resolved on a third baseline indicating the presence of warm dusty material surrounding FU~Ori out to several tens of AU (see figure~\ref{fig:quanz}).

Our mid-infrared (MIR) observations rule out one of the two disk models for FU~Ori previously presented by Malbet and collaborators\cite{2005A&A...437..627M} and we find that, quantitatively, the SED and the observed MIR visibilities can be fitted reasonably well with a simple analytical disk model prescribing a double power law for the effective temperature distribution of the protoplanetary disk. Within the innermost 3\,AU the temperature decreases with $T\propto R^{-0.75}$ farther out the temperature goes with $T\propto R^{-0.53}$. This indicates that in the innermost disk regions, the disk luminosity is accretion dominated, whereas further out it is irradiation dominated. This is expected for actively accreting objects, since the luminosity released by accretion is a much steeper function of distance to the star than that by absorption of radiation from the central star and the hot innermost disk regions ($L_{accretion} \propto R^{-4}$, whereas $L_{absorption} \propto R^{-2}$, roughly). The correlated flux indicates that 95\% of the 8-13\,$\mu$m flux comes from within the inner 25\,AU of the disk for the shortest of our baselines while 65\% of this flux arise from the inner 13\,AU for our longest baseline. The shapes and strengths of the total 8--13\,$\mu$m spectrum and the (spatially resolved) correlated spectra indicate that most dust particles within the accretion disk are amorphous and already significantly larger than typical particles in the ISM. No spectra, neither the total nor the spatially resolved correlated spectra bear significant traces of crystalline silicates. Given the high accretion rate and disk temperature of the system this is unexpected and requires further investigations. These results will appear shortly in print (Quanz et al. 2006, ApJ, in press).

\subsubsection{The eruption of V1647 Ori}
\label{sec:Abraham}

   \begin{figure}
   \begin{center}
   \begin{tabular}{c}
   \hspace{-0.5cm}
   \includegraphics[height=6.5cm]{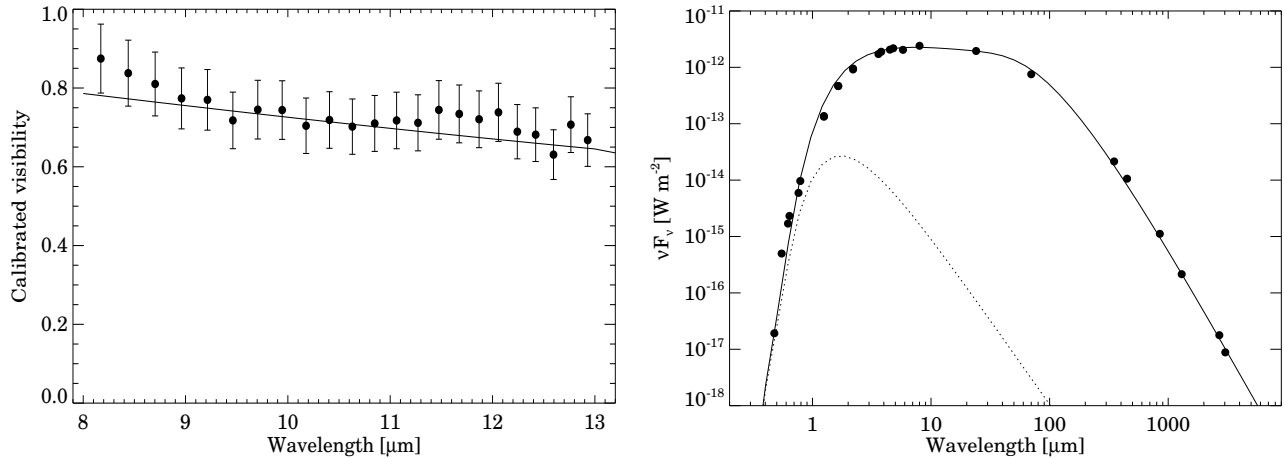}
   \end{tabular}
   \end{center}
   \caption[example]
   { \label{fig:abraham}
$Left$: Calibrated visibilities of V1647 Ori as a function of wavelength. The uniform error bars of 10\,\% reflect our conservative estimate of the uncertainties. The solid line represents the model visibility curve of V1647 Ori. $Right$:Spectral energy distribution of V1647~Ori. All optical and infrared data plotted were obtained in February-March 2004.  The thick line is the best fitting disk model (see text); the dotted line represents a black body approximation to the stellar photosphere.
}
   \end{figure}

In January 2004 a new reflection nebula (McNeil's Nebula) appeared in the LDN\,1640 dark cloud of the Orion\,B molecular cloud complex\cite{2004IAUC.8284....1M}. V1647\,Ori, whose $\sim$4\,mag outburst in the I-band caused the appearance of McNeil's Nebula, is a low-mass pre-main sequence object\cite{2004A&A...419L..39A}, which probably belongs to either the FU~Orionis or the EX~Lupi class of eruptive young stars. The object had been gradually fading until October 2005, when the eruption rapidly ended\cite{2005IBVS.5661....1K} .

V1647\,Ori was observed with MIDI on the UT3-UT4 baseline of the VLTI on March 2, 2005. The projected baseline length was 56\,m with a PA=112$^{\circ}$. From the measured visibilities (figure~\ref{fig:abraham}) we derive a linear size of the mid-infrared emitting region of $\sim$3\,AU at 8\,$\mu$m and $\sim$7\,AU at 13\,$\mu$m. The apparent increase in source size with wavelength indicates that the temperature of the emitting material radially decreases outward.
There are proposals that the FU Ori phenomenon is triggered by a close companion\cite{2004ApJ...608L..65R}. Such a companion would cause sinusoidal variations in the spectrally resolved visibilities with appropriate baseline position angles. The shape of our visibility curve suggests that no companion is present at the measured position angle whose separation is less than 100\,AU and contribution to the 10\,$\mu$m system flux is greater than 10\%. Also, the acquisitions images do not show any companions.

We found that a simple disk model is able to fit both the observed visibility values (solid line in left panel in figure~\ref{fig:abraham}) and the spectral energy distribution simultaneously (right panel), with the following model parameters: $T(1 \rm {AU})=680$\,K and $T{\propto}R^{-0.53}$, inner and outer disk radii 7\,R$_{\odot}$ and 100\,AU, respectively, surface density ${\Sigma}{\propto}R^{-1.5}$, disk mass $M_{d}=0.05 \rm {M}_{\odot}$, inclination angle $60^{\circ}$. This disk structure clearly differs from the canonical model usually proposed for FUORs (an optically thick, geometrically thin accretion disk with $T{\propto}R^{-0.75}$, \cite{1973A&A....24..337S}). Our model exhibits a shallower temperature profile of $T{\propto}R^{-0.53}$, whose exponent is similar to that of flared disk regions in standard circumstellar disk models. These results emphasize that eruptive young stars form a rather heterogeneous class, and many FUORs may require a different model than the canonical one. These results have appeared recently in print\cite{2006A&A...449L..13A}.

\subsection{The dust properties in T\,Tauri star disks}
\label{sec::Ttau_mineralogy}

In June 2003, several intermediate mass (HAE) stars were observed with MIDI and the spectral signatures of their dust were investigated. The main result was that \emph{the dust properties in the disks are not uniform}. Rather, a strong gradient was found in both the crystallinity and the average size of the dust grains as a function of distance to the central star. The grains in the innermost disk region are larger, and have a higher crystallinity than the grains at larger distances. These results are discussed in section~\ref{sec:van_boekel} and appeared in print\cite{2004Natur.432..479V}. 

\begin{figure}
   \begin{center}
   \begin{tabular}{cc}
   \includegraphics[height=5cm]{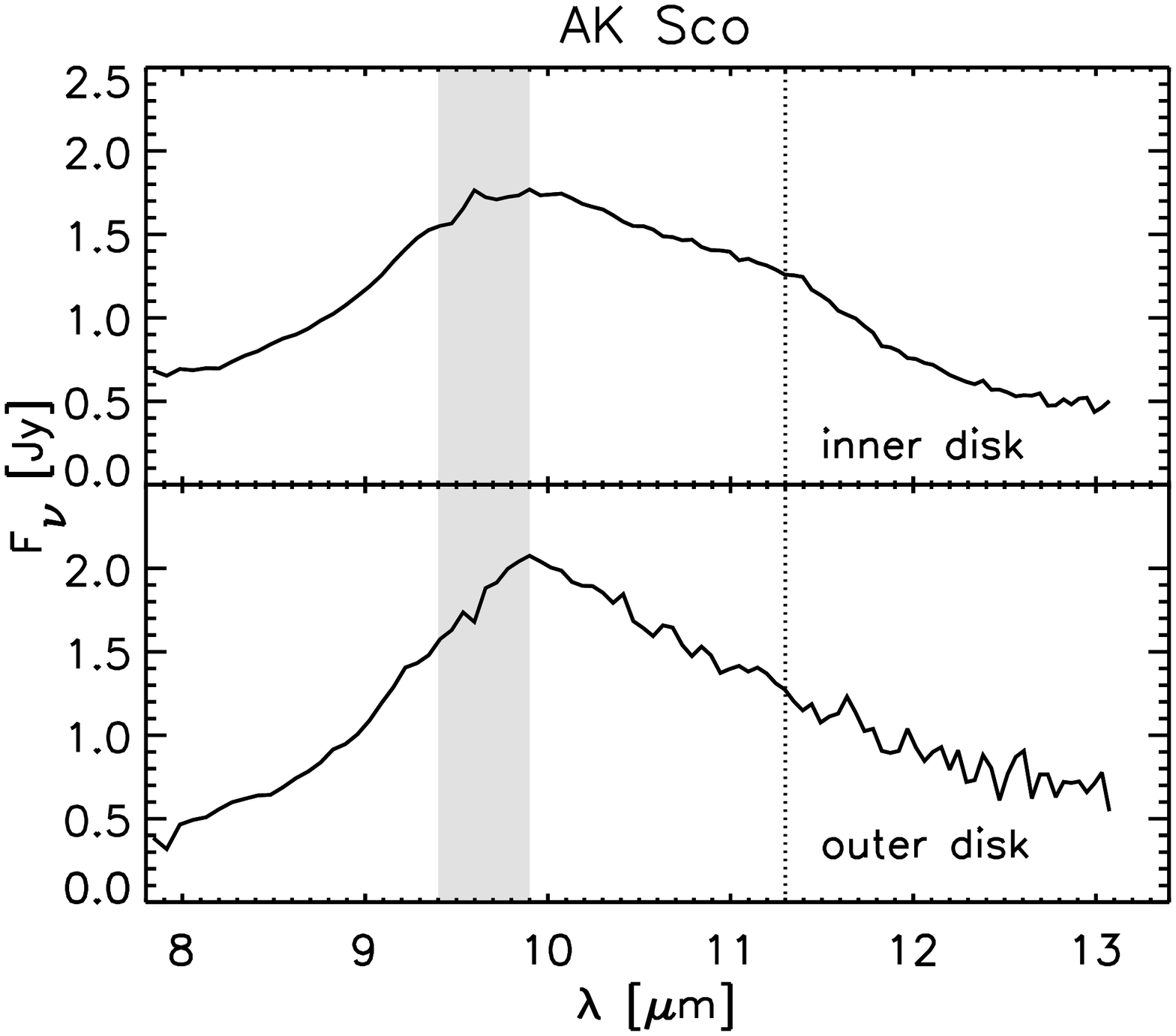} &
   \includegraphics[height=5cm]{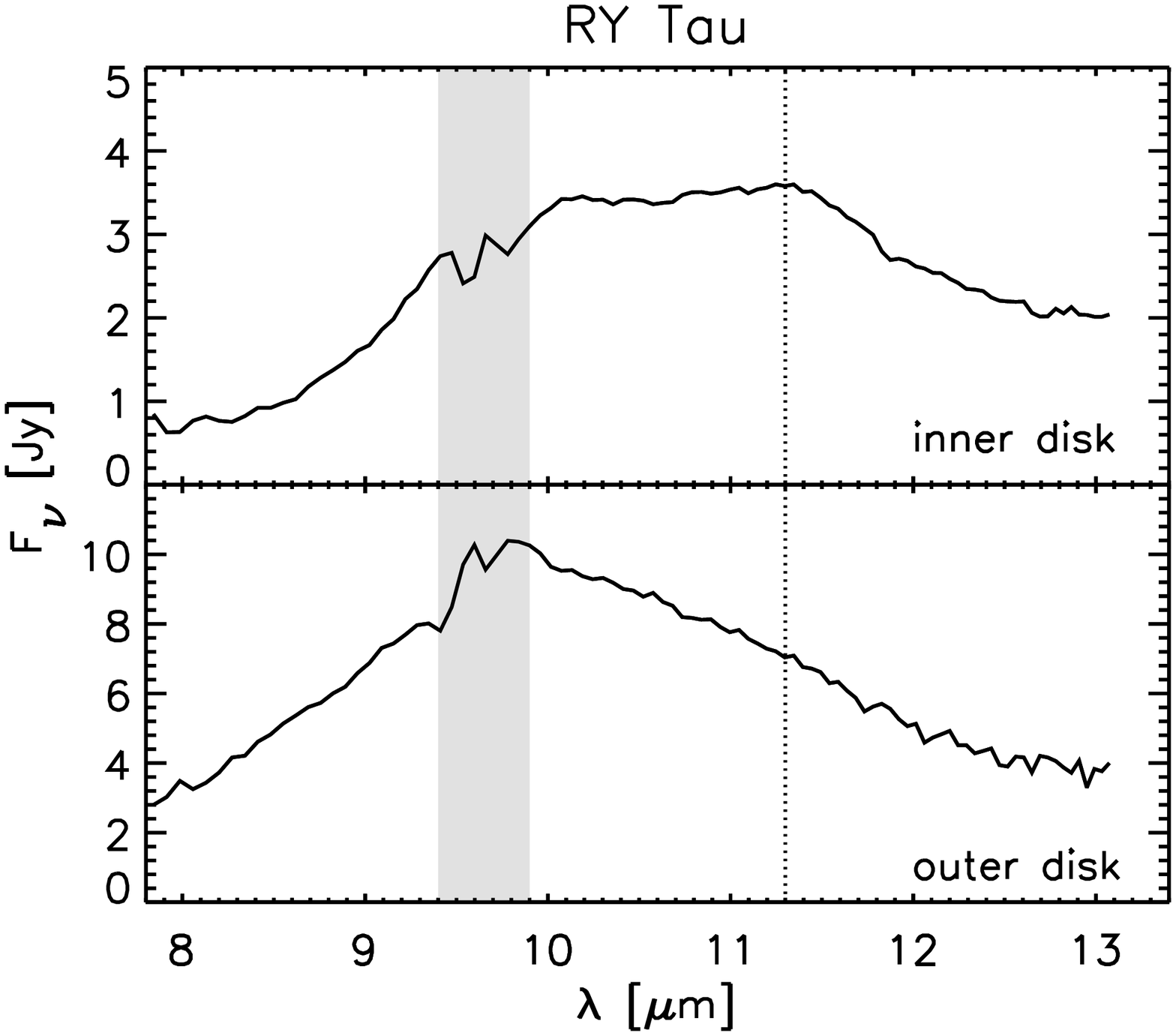}
   \end{tabular}
   \end{center}
   \caption[example]
   { \label{fig:Ttau_mineralogy}
The spatially resolved mineralogy of T\,Tauri star disks. The upper panels show the MIDI correlated spectra which are dominated by the innermost disk regions. The lower panels show the spectra of the outer disk regions. The dust in the inner regions is more evolved than further out (compare to figure~\ref{fig:processing}). The grey vertical bands indicate where the (terrestrial) Ozone band reduces the quality of the spectral calibration.
}
   \end{figure}

In figure~\ref{fig:Ttau_mineralogy} we show measurements very similar to the above mentioned ones. Here however, the targets are T\,Tauri stars. The upper spectra show the \emph{correlated} flux as measured by MIDI (see also figure~\ref{fig:144432_3baselines}), which is dominated by the innermost disk regions, here: ``inner disk''. In the lower spectra, the inner disk spectrum has been subtracted from the total flux, and thus we see the whole disk \emph{except} the innermost part, here: ``outer disk''. Comparing to the spectra in figure~\ref{fig:processing}, it is immediately apparent that the dust in the inner disk is more evolved than in the outer disk, though the effect is less pronounced than in the HAE stars. Nonetheless, the same trend that was seen in the HAE stars is also present in T\,Tauri stars, further confirming the similarities between the two. These results will appear in print (Leinert et al., in prep.).

\section{MIDI results: intermediate mass stars} 
\label{sec:intermediate_mass}

\subsection{The structure of HAE star disks}
\label{sec:HAE_disk_structure}
Here we present results of studies into the structure of Herbig~Ae star disks. We show both a study in which a sample of stars was measured on only one baseline (section~\ref{sec:leinert}), and two studies in which a single object was studied on a larger number of baselines (sections~\ref{sec:preibisch} and~\ref{sec:correia}).

\subsubsection{Flaring vs. self-shadowed disks}
\label{sec:leinert}

   \begin{figure}
   \begin{center}
   \begin{tabular}{c}
   \includegraphics[width=17.15cm,angle=0]{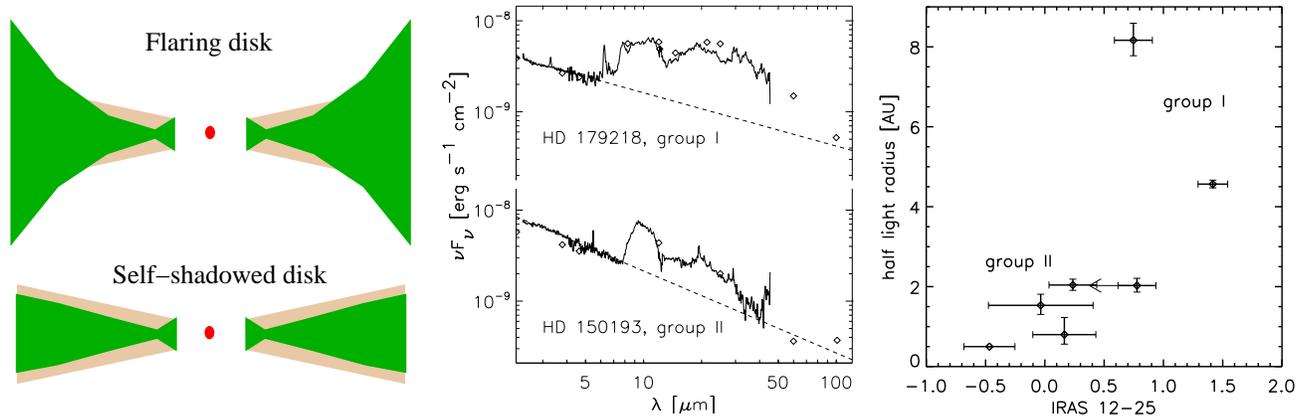}
   \end{tabular}
   \end{center}
   \caption[example]
   { \label{fig:leinert}
Sketches of disk geometries (``flaring'' and ``self-shadowed'', left panel), hypothesized to explain two types of SEDs observed in Herbig~Ae stars (group~I and group~II, middle panel). It is predicted that sources with group~I SEDs appear larger than group~II sources. Right panel: the spatial scale of the infrared emission as measured by MIDI at 12.5\,$\mu$m. Indeed, group~I sources appear significantly larger than group~II sources.}
   \end{figure}

Based on the infrared SED, it was first noted in ISO observations\cite{2001A&A...365..476M} that HAE stars can be divided into two main groups: ``group~I'' sources that have a very strong, rising IR excess peaking around 60\,$\mu$m, and ``group~II'' sources displaying a more moderate and less steeply rising IR excess (see middle panel of figure~\ref{fig:leinert}). The near infrared SED is very similar between both groups, suggesting a similar structure of the innermost (few) AU of the disk where the hottest dust is present. The cooler dust at larger radii is responsible for the emission at longer wavelengths. In the disk hypothesis, group~I sources have a ``flaring'' outer disk geometry, such that the disk can intercept and reprocess stellar radiation out to large distances from the star. In group~II sources the outer disk is not flaring, but ``self-shadowed'' instead (left panel of figure~\ref{fig:leinert}). The outer disk lies in the shadow cast by the ``puffed-up'' inner disk region, and does not directly ``see'' the central star. The lower dust temperature in group~II sources explains the reduced IR excess at longer wavelengths with to their group~I cousins. Furthermore, it predicts the infrared emission in group~II sources to be more compact.

Leinert and collaborators have observed a sample of HAE stars in June 2003 during MIDI's first large observing campaign\cite{2004A&A...423..537L}. The observations were performed at the 103\,m UT1-UT3 baseline and MIDI was operated in prism mode, yielding fringes at a spectral resolution of approximately 30. All sources were spatially resolved at 10\,$\mu$m, most for the first time. The measured visibility curves were compared to those predicted by fully independent model fits\cite{2003A&A...398..607D}, in which disk models\cite{2001ApJ...560..957D} were tuned to fit the SED only (see figure~1 and~4 of\cite{2004A&A...423..537L}). Good qualitative agreement is observed between the measured visibilities and those predicted solely on the basis of SED fits. In particular, the sources with a group~I SED show significantly lower visibilities than those with a group~II SED. 
In order to derive a simple estimate of the disk size, independent of the relatively sophisticated and complex models described above, a much simpler model was fit to the MIDI visibilities. This model consists of an optically thin distribution of grey dust particles with a power law surface density profile. The only fit parameter is the slope of density profile. We find steeper slopes for the group~II sources, indicating more compact emission than for the group~I sources. In the right panel of figure~\ref{fig:leinert} we plot the derived typical size of the emission, against an infrared color determined from IRAS photometry (group~II sources are blue and to the left, group~I sources are red and to the right). It is immediately apparent that the spatial extent of the 10\,$\mu$m emission in group~I sources is \emph{larger} than in group~II sources. MIDI has thus provided strong supportive evidence for the hypothesis that the former sources have flaring disks whereas the latter do not, by direct measurement on the relevant spatial scales.

As the first main result the existence of flaring and self-shadowed disks has been confirmed, but the full potential of MIDI has not yet been exploited. By comparing visibilities measured on several baselines of different lengths it will be possible to establish the presence of a puffed-up inner rim, and the ``shadowed region'' immediately behind the inner rim. The bright, ring-like emission expected from the inner rim and the relatively low intensity emission of the region just outward of the inner rim leave a typical imprint in the predicted visibilities (a ``plateau'' like visibility curve, see \cite{2005A&A...441..563V} for details). The needed baseline coverage and visibility accuracy for such studies are well matched to the capabilities of MIDI and observing campaigns to gather the required measurements are currently ongoing.


\subsubsection{The disk of HR 5999}
\label{sec:preibisch}

   \begin{figure}
   \begin{center}
   \begin{tabular}{c}
\hspace{-0.5cm}
   \includegraphics[height=8cm,angle=0]{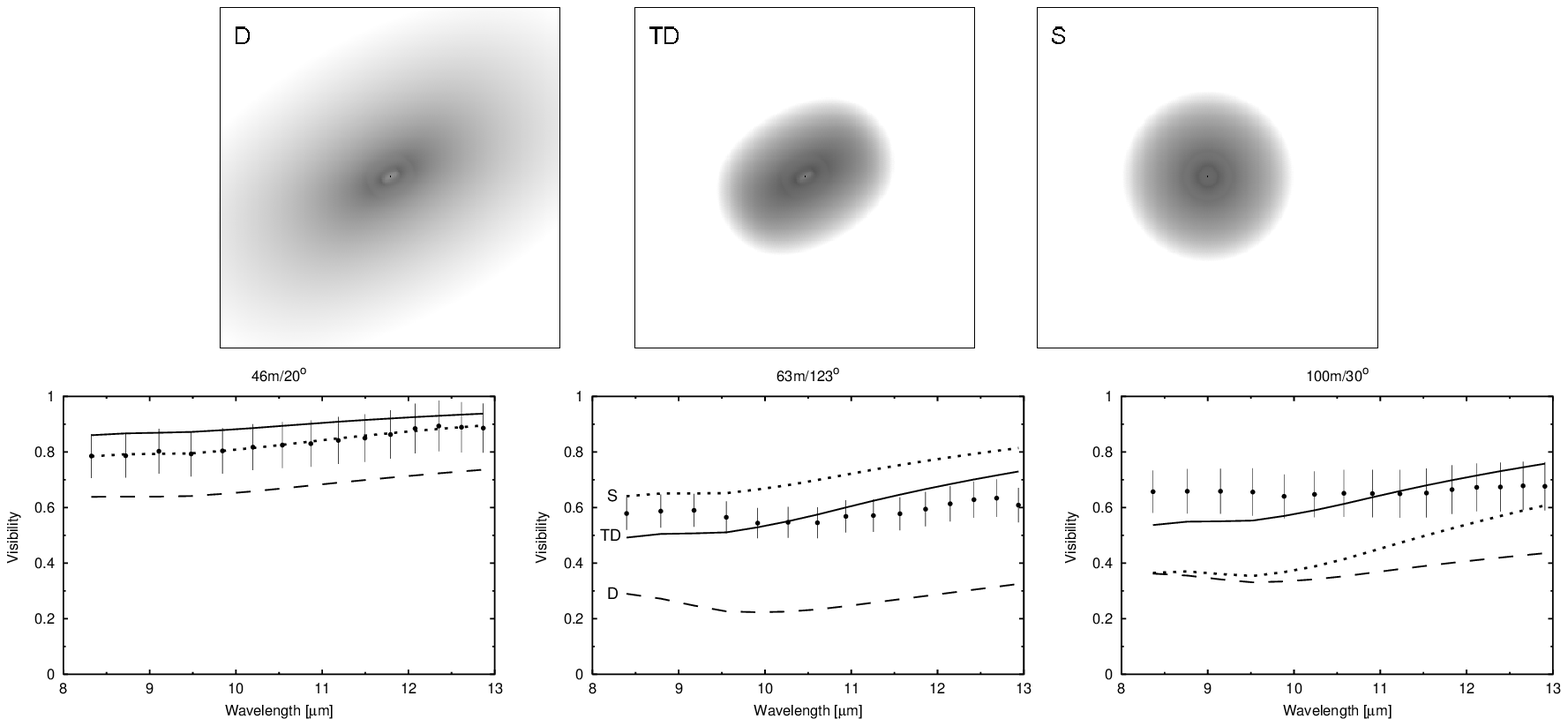}
   \end{tabular}
   \end{center}
   \caption[example]
   { \label{fig:preibisch}
Radiative transfer model images, model visibility curves and measured visibility curves for the Herbig Ae star HR\,5999. Shown are models for a disk (D, dashed lines in visibility plots), a truncated disk (TD, solid lines), and a spherical envelope (S, dotted lines).
}
   \end{figure}

HR\,5999 is a young ($\sim$0.5\,Myr), relatively massive ($\sim$3.2\,M$_{\odot}$) and luminous ($\sim$85\,L$_{\odot}$) Herbig Ae star at a distance of 210\,pc\cite{1998A&A...330..145V}. It has a very ``blue'' spectral energy distribution with relatively weak far-infrared emission (i.e. it is a group~II source, see section~\ref{sec:leinert}). It was observed with MIDI in April and June 2004 on the UT2-UT3 and UT1-UT3 baselines, a total of 10 visibility points were obtained at projected baseline lengths between 39 and 102\,m, at position angles ranging 15 and 173 degrees E of N. The derived visibility values between $\sim$0.5 and $\sim$0.9 show that the mid-infrared emission from HR\,5999 is spatially resolved but very compact. The characteristic size of the emission region depends on the projected baseline length and position angle, and ranges between $\sim 5-15$~milli-arcseconds (Gauss FWHM), corresponding to physical sizes of $\sim$1--3\,AU. To derive constraints on the geometrical distribution of the dust we compare our interferometric measurements to 2D, frequency-dependent radiation transfer simulations of circumstellar disks and envelopes. In figure~\ref{fig:preibisch} we show images and visibility curves of the best fit models of a disk, a truncated disk and a spherical halo, along with the MIDI data on 3 of the baselines. Models of spherical envelopes do not reproduce the observed degree of asymmetry present in the MIDI data. For disk models with radial power-law density distributions, the relatively weak but very extended emission from outer disk regions ($>$3\,AU) leads to model visibilities that are significantly lower than the observed visibilities; these models are thus inconsistent with the MIDI data. Disk models in which the density is truncated at outer radii of $\sim$2--3\,AU, on the other hand, provide good agreement with the data. A satisfactory fit to the observed MIDI visibilities of HR\,5999 is found with a model of a geometrically thick disk, which is truncated at $2.7$\,AU and seen under an inclination angle of $60^{\circ}$ (i.e.~closer to an edge-on view than to a face-on view). Models of a geometrically thin disk seen nearly edge-on cannot achieve agreement between the observed and predicted visibilities. The reason why the disk is so compact remains unclear; we speculate that it has been truncated by an unseen close binary companion. An alternative explanation for the small measured size may be very strong shadowing of the outer disk by the puffed-up inner rim, leading to low temperatures and thus fainter 10\,$\mu$m emission from the outer disk than in our model. These results will appear shortly in print (Preibisch et al. 2006, A\&A, in press).

\subsubsection{R~CrA: disk vs. nebula orientation}
\label{sec:correia}

   \begin{figure}
   \begin{center}
   \begin{tabular}{cc}
\hspace{-0.4cm}
   \includegraphics[height=5.8cm]{R_CrA-rim-sed.ps} &
   \includegraphics[height=6.1cm]{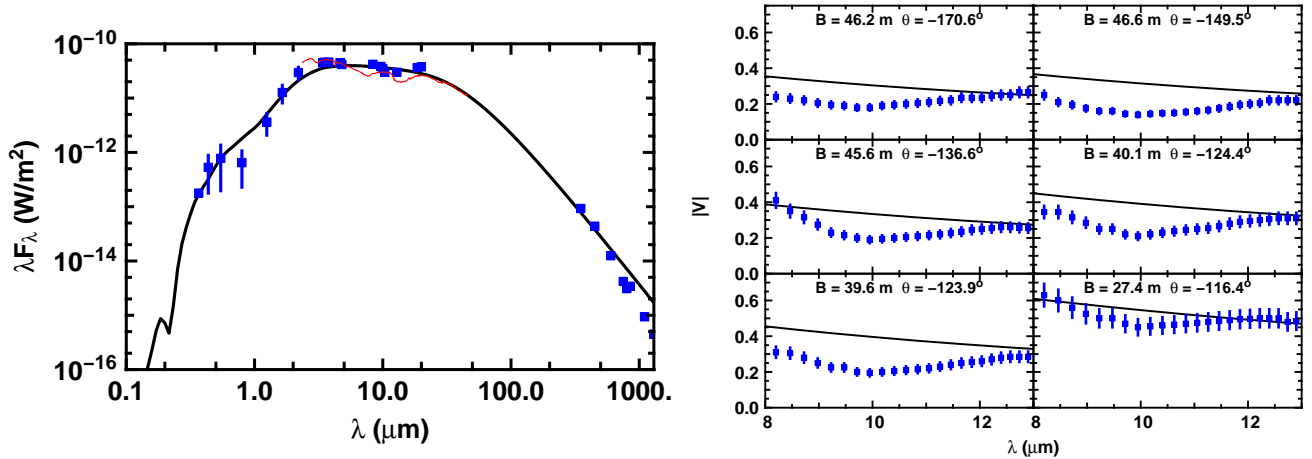}
   \end{tabular}
   \end{center}
   \caption[example]
   { \label{fig:zinnecker}
Simultaneous fit of the R~CrA SED (left) and visibilities (right). The best fit model shown is composed of the star ($R$=1.8\,R$_\odot$, $T_{\mathrm{eff}}$=7500\,K), an accretion disk (power-law temperature profile index 0.55, $T$(1\,AU)=650\,K, minimum radius 0.16\,AU, maximum radius 30\,AU, inclination 30$^{\circ}$) and a ring representing the puffed-up inner rim (radius 0.12\,AU, width 0.05\,AU, temperature 1500\,K, same inclination). We assumed $A_{\mathrm{v}}$=3\,mag and a distance of 130\,pc. Squares are measurements, while the continuous line is the best fit model. In the SED plot, the ISO spectrum is also shown (grey line).
}
   \end{figure}

R~CrA is a young Herbig Ae star located at the center of the small ``Coronet'' cluster\cite{1984MNRAS.209P...5T} at a distance of 130\,pc\cite{1999AJ....117..354D}.
It exhibits several of the typical characteristics indicating the presence of a circumstellar disk around a young star, in particular a large IR-excess\cite{2004A&A...426..151A}, a high degree (8\%) of optical linear polarization, 
\cite{1987ApJ...317..231B,2000MNRAS.319..337C} and optical brightness variations of the UX Ori type\cite{2003ApJ...594L..47D}.

We observed R~CrA with MIDI at the VLTI in July 2004 and obtained 6 sets of spectrally dispersed visibilities in the spectral range 8--13\,$\mu$m with projected baselines lengths ranging from 27 to 47\,m, and position angles between 9 and 63~degrees. A first comparison of the data with a simple geometrical model of an inclined ring shows that the emitting region has a typical size of 6--10 AU, consistent with previous MIDI observations of Herbig Ae stars with disks\cite{2004A&A...423..537L}. The inclination is constrained to 44$^{+8}_{-17}$ degrees, roughly perpendicular to the symmetry axis of a bipolar reflection nebula derived from NIR imaging polarimetry\cite{2000MNRAS.319..337C}. Using additional baselines with a larger range of position angles, we are able to derive a disk semi-major axis at position angle 59$^{+9}_{-8}$ degrees E of N, which is also perpendicular to the symmetry axis of the nebula.

Figure~\ref{fig:zinnecker} shows our first attempt to model both SED and mid-IR visibilities with a simple model composed of the star ($L$=10\,L$_{\odot}$), a bright ring at 25\,R$_{\odot}$ ($\sim$0.1\,AU) and a disk emitting locally as a blackbody, self-shadowed from 25 to 35\,R$_{\odot}$ and emitting from 35\,R$_{\odot}$ to 30\,AU. The bright ring represents the ``puffed-up'' inner rim of the disk. It is encouraging to see that both SED and visibilities are relatively well reproduced by such a model. In particular, a robust temperature profile slope of $\sim$\,0.5 (i.e. $T \propto R^{-0.5}$), characteristic of a passive disk, is deduced by this simultaneous SED-visibility modeling. The more ``curved'' shape of the observed visibilities with respect to the modeled one may be due to the increased dust opacity in the silicate feature\cite{2005A&A...441..563V}, which is not included in our model. More detailed modeling is necessary to further constrain the disk structure including a possible contribution from an envelope.

\subsection{The dust properties in HAE star disks}
\label{sec:van_boekel}

The rich spectral signatures of small silicate grains at infrared wavelengths allow the study of the dust properties in the surface layers of circumstellar disks. In the spectra of many disks, signatures of dust processing (grain growth and/or crystallization) are seen\cite{2001A&A...375..950B}. Contrary to existing single telescopes, MIDI is able to \emph{spatially resolve} the infrared emission. Thus, we can determine \emph{where} in the disk the grown and crystallized grains are present. Crystalline silicates, which may be formed in the hot innermost disk regions, are found also at lower temperatures, i.e. larger distances from the star\cite{1998A&A...332L..25M}. \emph{If} these crystals are transported to large distances by radial mixing, then the abundance of crystalline silicates as a function of distance to the star may provide a direct probe of the mixing process.

\begin{figure}
   \begin{center}
   \begin{tabular}{c}
   \includegraphics[height=8cm]{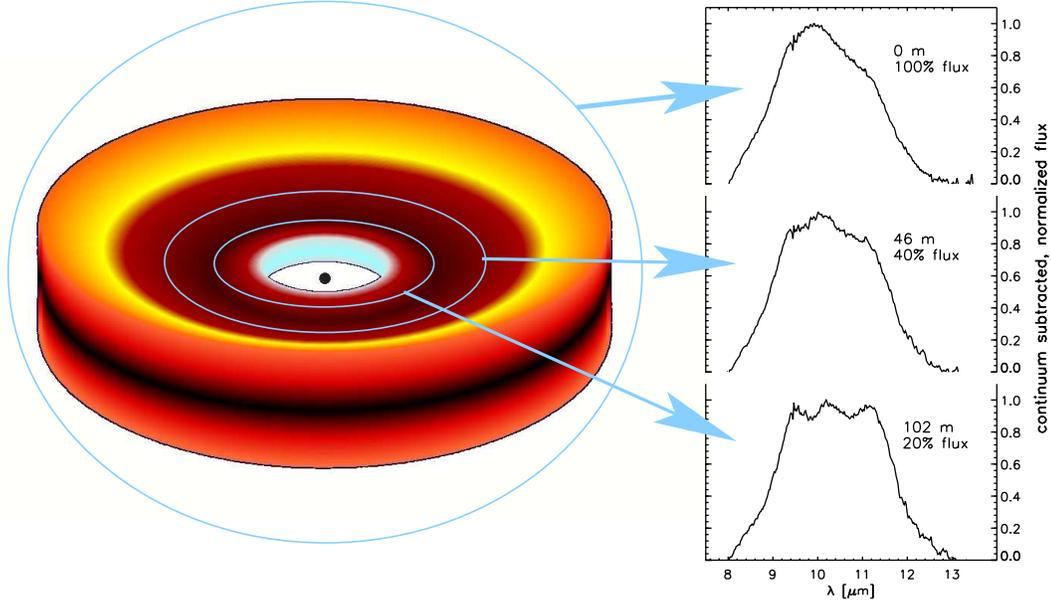}
   \end{tabular}
   \end{center}
   \caption[example]
   { \label{fig:144432_3baselines}
The 10\,$\mu$m silicate feature of HAE star HD\,144432 on three different spatial scales. The regions that pay the most important contribution to the different spectra have been indicated. The integrated spectrum in which all light contributes (top right) shows mostly pristine (small, amorphous) grains. The interferometric spectrum taken with MIDI on a baseline of 102\,m (lower right\cite{2004Natur.432..479V}) is dominated by roughly the central AU of the disk, and shows highly evolved dust, with larger grains, high crystallinity (compare to figure~\ref{fig:processing}). The MIDI spectrum at a baseline of 46\,m (middle right) probes a larger region than the 102\,m spectrum, and already has a significant contribution from pristine dust. Note that the interferometric point spread function is not ``round'' as sketched here. Rather, it has a cosine behavior with period $\lambda/B$, where $\lambda$ is the observing wavelength and $B$ the Baseline. However, assuming that the emission of the disk on larger scales is smooth (i.e. does not have small scale features that contribute significantly to the emission), this emission is ``resolved out'' by the interferometer. Only in the center, where the high dust temperatures cause a large flux from a small region, we get a significant contribution to the interferometric spectra. Disk sketch courtesy of Vincent Icke.
}
   \end{figure}

In figure~\ref{fig:144432_3baselines} we show three N-band spectra of HD\,144432. The upper spectrum is the ``normal'' spectrum taken by a single telescope (one may consider it the correlated spectrum at a baseline of 0\,m). Here, the source is spatially unresolved and we see the light from the entire disk region that is warm enough to emit at 10\,$\mu$m, i.e. roughly the central 10-20\,AU. Comparing to the spectra in figure~\ref{fig:processing} we see that most of the material visible here is pristine, but the small ``shoulder'' at 11.3\,$\mu$m indicates that there is some processed material \emph{somewhere} in the system. The middle and lower plot show the correlated spectrum as measured by MIDI on baselines of 46 and 102\,m. The emission seen here is dominated by the disk regions within 3 and 1.5\,AU of the central star, respectively. The correlated spectrum at a baseline of 102\,m, which probes the smallest spatial scales, is very similar to the ``evolved'' spectrum in figure~\ref{fig:processing}. We made compositional fits to these spectra, including both amorphous and crystalline silicates in grain sizes of 0.1 and 1.5\,$\mu$m as dust species, in order to study both grain growth and crystallization. We find that the fraction of crystalline silicates is approximately 5, 12 and 36 percent for the spectra at baselines of 0, 46 and 102\,m, respectively. The fraction of material contained in large grains is 40, 86 and 93 percent by mass. Thus a clear trend is seen, indicative of both crystallinity and average grain size in the surface layer of the disk decreasing with distance to the central star.

Similar results were obtained for two additional objects (HD\,163296 and HD\,142527). All three disks display a much higher degree of processing in their innermost regions than further out. In the case of HD\,142527, there are crystalline silicates present also at larger radii, indicative of radial mixing of material. These results have appeared in print\cite{2004Natur.432..479V}. The combination of accurate visibility measurements on multiple baselines of different length will show whether or not the abundance of crystalline silicates changes with distance to the star in a way consistent with radial mixing.

\section{MIDI results: high mass stars} 
\label{sec:high_mass}

   \begin{figure}
   \begin{center}
   \begin{tabular}{c}
   \includegraphics[height=17.15cm,width=10.5cm,angle=90]{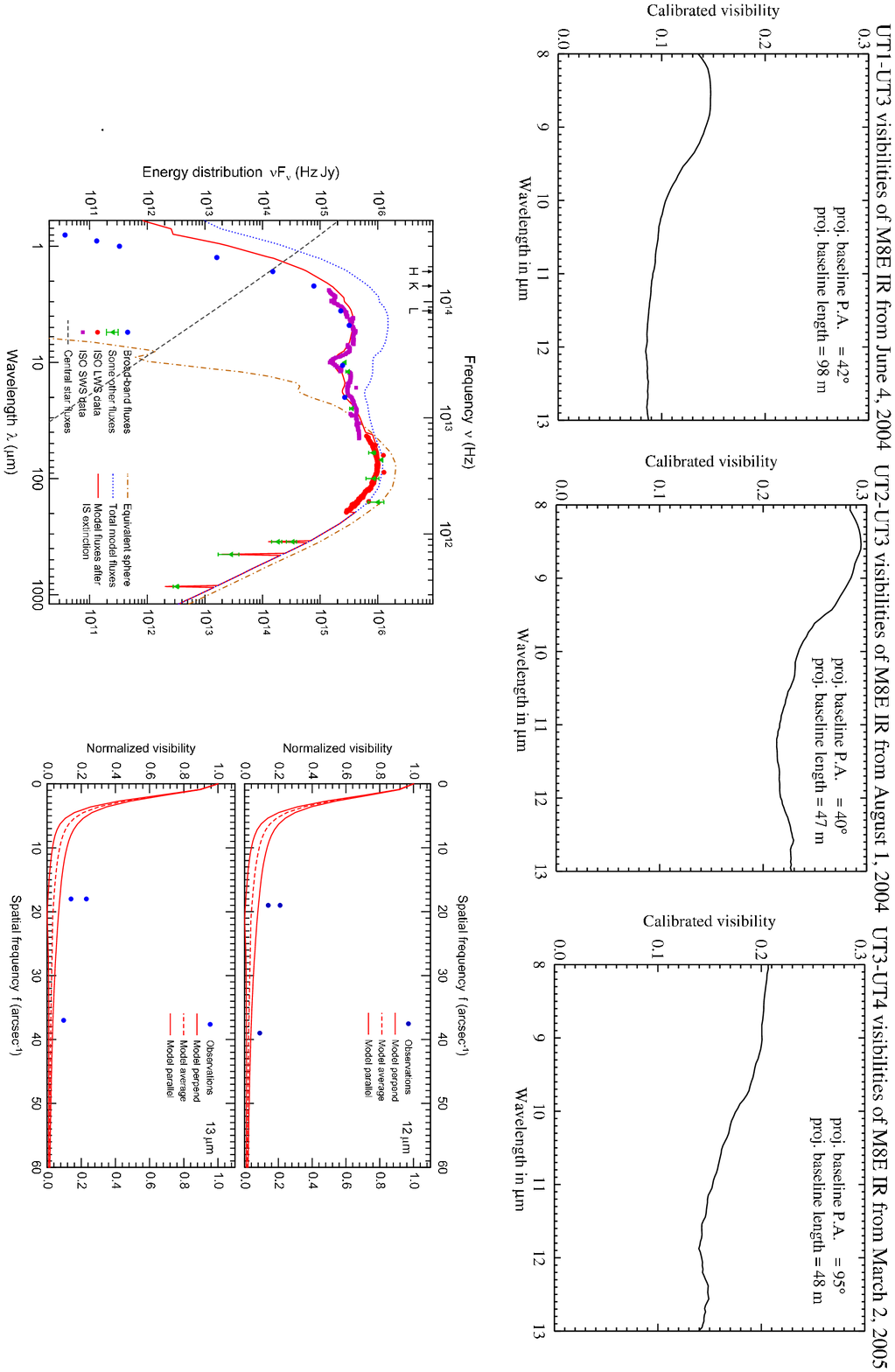}
   \end{tabular}
   \end{center}
\vspace{-0.3cm}
   \caption[example]
   { \label{fig:linz}
"Observational and modeling data for the massive YSO M8E--IR. {\it Upper panels}: Calibrated visibilities for three different baseline configurations. Note the relatively low visibility values for all three measurements which indicates that we clearly resolved the source (visibility error bars are typically 10\.\%). {\it Lower left}: Comparison of the observed SED for  M8E--IR and the predictions of our first radiative transfer model. {\it Lower right}: Comparison of observed and model visibilities at 12 and 13~$\mu$m. Modeled visibilities are still too low, while the difference in visibilities between the two viewing angles is about correct.
}
   \end{figure}

Massive stars that are currently being formed are very rare. This is due to the fact that, compared to lower mass objects, nature produces relatively few massive stars, and their formation time scales are very brief (of order 10$^5$\,yrs). Consequently, the regions where active high--mass star formation can be studied are located at large distances (typically 3 to 7~kpc). Massive stars are deeply embedded in their maternal clouds during their entire formation process. The very high extinction due to obscuring dust -- typically many tens of magnitudes -- causes young massive stars to be essentially invisible at optical and often even near-infrared wavelengths. Massive stars tend to form in highly clustered environments, leading to confusion problems in long wavelength observations that have poor spatial resolution. Due to all of the above mentioned reasons, observational studies of forming high mass stars are much more difficult than those of their low mass cousins, and our knowledge of high mass star formation is less detailed. A promising way forward is to perform high-spatial resolution observations at mid-infrared (MIR, 10--20\,$\mu$m) wavelengths. The high spatial resolution is needed to spatially resolve the thermal dust emission despite the large distances towards high-mass star forming regions. In the mid-infrared, the extinction due to obscuring dust is much lower than at shorter wavelengths, whereas the spectral energy distribution of high--mass young stellar objects exhibits a strong increase. The MIDI instrument provides the highest spatial resolution at MIR wavelengths available world-wide now and in the foreseeable future, making it an ideal instrument for the study of massive star formation.

There is growing observational evidence that massive star formation proceeds along a modified accretion scenario. This is mainly based on the presence of massive molecular outflows and also relatively well collimated jets\cite{2002A&A...383..892B}. Circumstellar disks may be a way to circumvent the problems of radiation pressure when accreting mass on massive YSO\cite{2002ApJ...569..846Y}. However, such disks turn out to be elusive observationally.
Attempts in the near--infrared, for instance by means of AO--assisted direct imaging or CO band-head spectroscopy, have limited success, are affected by still large extinction, and only rarely approach the O-star regime\cite{2004A&A...427L..13B}. Thermal infrared spectroscopy of such molecular tracers might be a future possibility to catch more embedded objects -- this is also a science driver for the 2$^{nd}$ generation VLTI project \emph{Matisse} (see section~\ref{sec:matisse}).

Previous massive disk investigations mostly apply (sub-)mm interferometry techniques. Thereby, the achievable spatial resolution slowly approaches the range necessary for clear disk detections, but apart from the VLA 7-mm system, sub-arcsecond resolution in the millimeter range is still difficult to attain. Several studies reveal relatively large circumstellar structures\cite{2004Natur.429..155C , 2002ApJ...566..982Z} that might turn out to be flattened massive envelopes or tori of several 1000\,AU in size. However, the pivotal question is whether real accretion disks around massive YSOs exist that actually feed the forming massive stars in order to build up their mass. MIDI is used to probe the warm circumstellar dust on scales of 20 to 100~AU.

\newpage

Massive stars that are bright in the MIR form a natural choice of targets. To date, we obtained fringes for 8 sources. The objects are mostly members of the class of so--called BN-type objects\cite{1990A&A...227..542H} after the prototypical Becklin--Neugebauer object in Orion. The targets are compact and appear unresolved in previous MIR observations with 4-m class telescopes, but the nearly-diffraction limited MIDI acquisition\footnote{MIDI can also make ``normal'' single telescope images with a field of view of 2 arcseconds. This is not an interferometric measurement, even though the light follows the whole path from the telescopes, through the delay lines and into the interferometric laboratory.} images hint that some are marginally resolved by a single 8\,m telescope at 10\,$\mu$m. 

With regard to our disk research with MIDI, the object M8E--IR is of particular interest. Based on a lunar occultation study at 3.8 and 10 $\mu$m, Simon and collaborators\cite{1985ApJ...298..328S} presented a two--component model for this massive YSO at a distance of 1.5 to 1.8~kpc with a luminosity of roughly 2.5$\times$$10^4$ L$_{\odot}$\cite{1984ApJ...278..170S}. They propose the existence of a large, spherically symmetric component 100\,mas in diameter and, of a smaller, elongated component (FWHM 6--21\,mas). The latter was suggested to trace a flared circumstellar disk seen nearly edge--on, with the major axis at a position angle of roughly 150$^\circ$. However, subsequent CO molecular line studies\cite{1992ApJ...386..604M} reveal a large--scale bipolar molecular outflow with roughly the same orientation (145$^\circ$). A disk around the driving source M8E--IR would be expected to be more or less perpendicular to the flow. Thus, already L\"owe and collaborators\cite{1997svlt.work..379L} have speculated that the structure ``seen'' by Simon et al. actually delineates the inner parts of the outflow cone.

We have observed M8E--IR with MIDI in combination with the Unit Telescopes at 7 different (baseline, position angle) configurations with projected baseline lengths ranging from 43 to 97\,m. The object is clearly resolved in all our configurations with visibilities between 0.09 and 0.35. If we assume a Gaussian intensity distribution of the source, the visibilities indicate a FWHM of $\sim$20--25\,mas (8.5\,$\mu$m) and 32-38\,mas (12.0\,$\mu$m), which is in rough agreement with the extension of the small component of Simon et al.\cite{1985ApJ...298..328S} (Remember that 30\,mas correspond to 54\,AU at 1.8\,kpc). Indeed, the sizes seem to be 15--20\,\% smaller for the baseline with P.A.=138$^\circ$, i.e. roughly along the large--scale molecular outflow. However, the whole set of observations cannot be straightforwardly interpreted in terms of a simple disk perpendicular to that flow. Probably, the small--scale geometry is more complex, and even the warm inner parts of the outflow cone might contribute significantly to the interferometric signal. We furthermore mention that a Gaussian ansatz is a poor match for the intensity distribution of a circumstellar disks and for that of low--visibility objects in general. Hence, we started more elaborate modeling by utilizing a 1.5\,D radiative transfer code\cite{1997A&A...318..879M} that models a flared disk + outflow cavities + envelope. Standard dust is assumed for the envelope while larger grains (radius $>$1\,$\mu$m) were used for the disk in order to also fit the SED of M8E--IR. Still, the agreement between observations and the modeled visibilities (based on the first 3 of the 7 measurements) is less than perfect (figure~\ref{fig:linz}). The assumed model yet results in too low visibilities. Both the model and the data reduction need further refinement.
After inclusion of all 7 measurements into the modeling and re-checking of the robustness of the data reduction, we can more reliably say whether the insinuated elongation on scales of 50\,AU is real and whether a disk + envelope model can convincingly reproduce the observations.

\section{A look ahead}
\label{sec:matisse}

   \begin{figure}
   \begin{center}
   \begin{tabular}{cccc}
   \includegraphics[height=5cm]{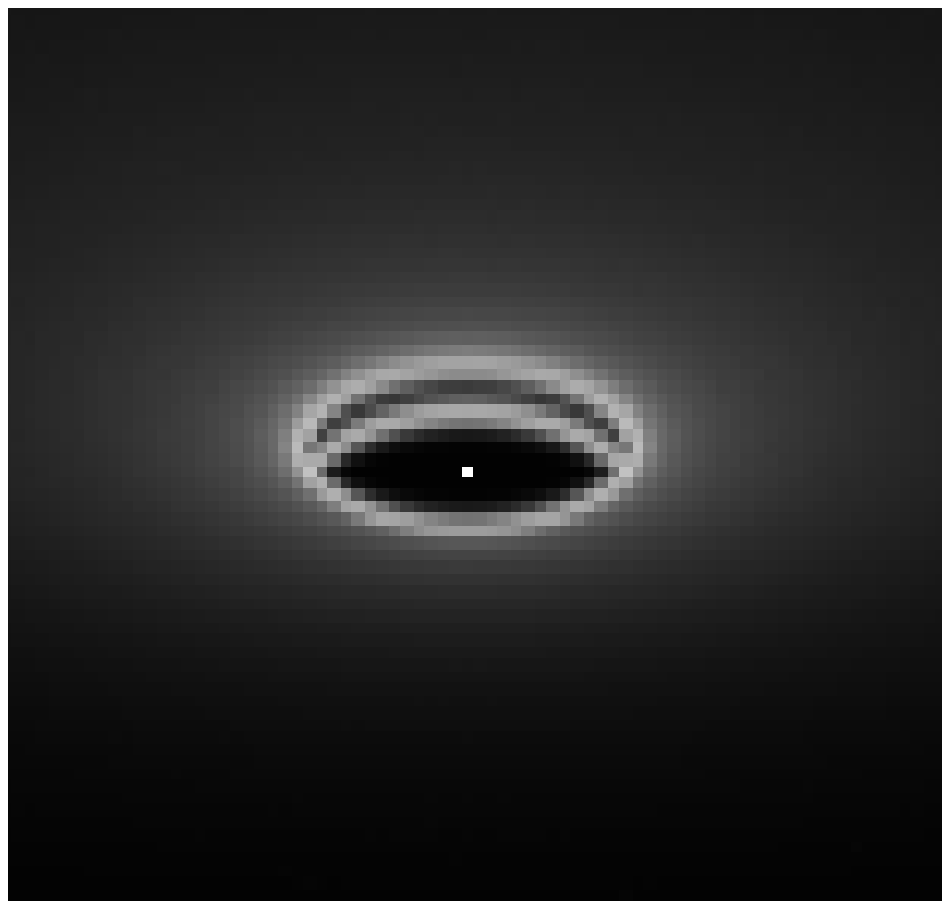} &
&
&
   \includegraphics[height=5cm]{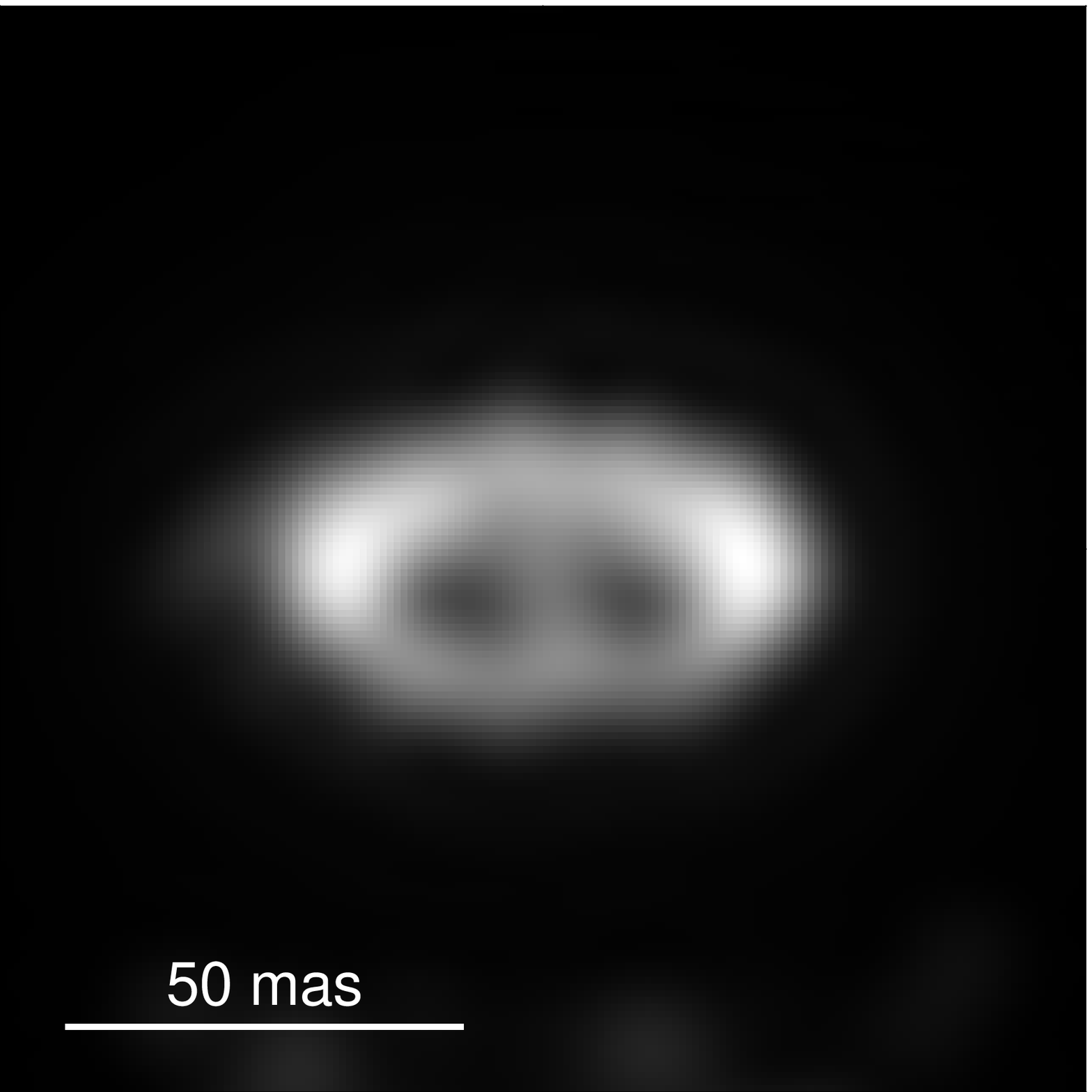}
   \end{tabular}
   \end{center}
   \caption[example]
   { \label{fig:matisse}
\emph{Left:} Simulated  10\,$\mu$m image of the inner region of a T\,Tauri circumstellar disk with cleared inner region, seen under an inclination of 60$^{\circ}$. A distance of 140\,pc is assumed, the size of the inner gap is 4\,AU. \emph{Right:} \emph{Matisse} image reconstructed from a data set that is roughly equivalent to 3 nights of observations with 4 ATs. The data set consists of 210 visibility amplitudes and 70 closure phase relations, a noise of 5\,\% in the (squared) visibilities and 10\,\% in the closure phases is assumed.
}
   \end{figure}

MIDI has for the first time allowed spatially resolved studies of circumstellar disks on the scales where accretion, dust processing and planet formation take place. The first measurements were in good qualitative agreement with theoretical predictions from the latest disk models, specifically it was shown that the hypothesized correlation between disk geometry and spectral energy distribution is indeed physical reality\cite{2004A&A...423..537L}. Furthermore it has been shown that the mineralogy in disks is a strong function of distance to the central star\cite{2004Natur.432..479V}. Oncoming MIDI observations providing more accurate visibilities on a larger number of different baselines will yield much tighter constraints on the disk geometry and on the distribution of the different minerals. However, like any instrument also MIDI has its limitations. As a two-element interferometer MIDI cannot measure the visibility phase which is needed for image reconstruction.  Ambiguity is unavoidable when studying objects that are not point-symmetric with respect to the center. Also, getting measurements on many baselines ("filling the uv-plane'') is very inefficient if only two telescopes at a time are used.

\newpage
For the second generation VLTI instrumentation, an instrument capable of combining 4 beams is currently under consideration\cite{2006MAT.....1....1W}. The Multi AperTure Mid-Infrared SpectroScopic Experiment (\emph{Matisse}) will be capable of measuring visibility amplitudes on 6 different baselines (quasi-) simultaneously, as well as interferometric (closure) phases. This allows for true image reconstruction, roughly at the level of complexity of VLBI observations in the radio domain (see figure~\ref{fig:matisse}). 
Furthermore, \emph{Matisse} will be capable of measuring in the L, M, N and Q bands (around 3.5, 4.8, 10 and 20~micron, respectively). This allows the study of dust of both higher and lower temperatures than the dust to which MIDI is most sensitive. With high spectral resolution in the L and M bands, gas emission lines can be used to probe the physical conditions throughout the inner disk regions in detail, this is a completely new capability compared to MIDI.

%
%

\acknowledgments     
It is a pleasure to thank all those involved in designing, building, commissioning and operating the VLT Interferometer and MIDI.

\bibliography{vboekel_spie06}   
\bibliographystyle{spiebib}   

\end{document}